%% file: ms.tex
\documentclass{article} % For LaTeX2e
\usepackage{iclr2019_conference,times}

\input{math_commands.tex}

\usepackage{hyperref}
\usepackage{url}
\usepackage{graphicx}
\usepackage{amsmath}
\usepackage{algorithm}
\usepackage[noend]{algpseudocode}
\usepackage{subfig}
\usepackage{python}
\usepackage{natbib}
\usepackage{float}
\usepackage{ctable}
\usepackage{caption}
\makeatletter
\def\BState{\State\hskip-\ALG@thistlm}
\makeatother

\title{Design of Intentional Backdoors in Sequential Models}

\author{Zhaoyuan Yang, Naresh Iyer\thanks{Authors contributed equally to this work}, Johan Reimann\footnotemark[1], Nurali Virani\footnotemark[1] \\
GE Research\\
Niskayuna, NY 12309, USA \\
\texttt{\{zhaoyuan.yang,iyerna,reimann,nurali.virani\}@ge.com} \\
}

\begin{document}

\maketitle

\begin{abstract}
Recent work has demonstrated robust mechanisms by which attacks can be orchestrated on machine learning models. In contrast to adversarial examples, backdoor or trojan attacks embed surgically modified samples with targeted labels in the model training process to cause the targeted model to learn to misclassify chosen samples in the presence of specific triggers, while keeping the model performance stable across other nominal samples. However, current published research on trojan attacks mainly focuses on classification problems, which ignores sequential dependency between inputs. In this paper, we propose methods to discreetly introduce and exploit novel backdoor attacks within a sequential decision-making agent, such as a reinforcement learning agent, by training multiple benign and malicious policies within a single long short-term memory (LSTM) network. We demonstrate the effectiveness as well as the damaging impact of such attacks through initial outcomes generated from our approach, employed on grid-world environments. We also provide evidence as well as intuition on how the trojan trigger and malicious policy is activated. Challenges with network size and unintentional triggers are identified and analogies with adversarial examples are also discussed. In the end, we propose potential approaches to defend against or serve as early detection for such attacks. Results of our work can also be extended to many applications of LSTM and recurrent networks. 
\end{abstract}

\section{Introduction}

Current research has demonstrated different categories of attacks on neural networks and other supervised learning approaches. Majority of them can be categorized as: (1) inference-time attacks, which add adversarial perturbations digitally or patches physically to the test samples and make the model misclassify them~\cite{fgsm, IPNN} or (2) poisoning attacks, which, on the other hand, corrupt training data and in case of trojans, embed carefully designed samples in the model training process to cause the model to learn incorrectly with regard to only those samples, while keeping the training performance of the model stable across other nominal samples~\cite{badnets, purdueTrojan}. The focus of this paper is on trojan attacks. In these attacks, the adversary designs appropriate triggers that can be used to elicit unexpected and unanticipated behavior from a seemingly honest model. As demonstrated in~\cite{badnets}, such triggers can lead to dangerous behaviors by artificial intelligence (AI) systems like autonomous cars by deliberately misleading their perception modules into classifying ‘Stop’ signs as ‘Speed Limit’ signs. 

Most research on trojan attacks in AI mainly focuses on classification problems, where model's performance is affected only in the instant when a trojan trigger is present. In this work, we bring to light a new trojan threat in which a trigger needs to only appear for a very short period and it can affect the model's performance even without the need to reappear in model's inputs later. For example, the adversary needs to only present the trigger in one frame of an autonomous vehicle's sensor inputs and the behavior of the vehicle can be made to change permanently from thereon. Specifically, we utilize a sequential decision-making formulation for the design of this type of threat and we conjecture that this threat applies to many applications of LSTM networks and is potentially more damaging in impact.

This work extends existing knowledge by providing: (1) a threat model and formulation for a new type of trojan attack for LSTM networks and sequential decision-making agents, (2) implementation and experimental results illustrating the threat, and (3) analysis of models with the threat and potential defense mechanisms. 

The organization of the paper is as follow: Section~\ref{sec:related} provides examples of related work. Section~\ref{sec:back} provides relevant background on deep reinforcement learning and LSTM networks. Section~\ref{sec:threat} describes the threat model in greater detail. Section~\ref{sec:implement} shows the implementation details, algorithms, experimental results, and identifies analogies with other adversarial attacks. Section~\ref{sec:def} provides examples of approaches for defending against such attacks. Finally, Section~\ref{sec:conc} provides conclusion and outlines some key directions for future research. 

\section{Related Work}
\label{sec:related}

Adversarial attacks on neural networks have received increasing attention after neural networks were found to be vulnerable to adversarial perturbations~\cite{IPNN}. Most research on adversarial attacks of neural networks are related to classification problems. To be specific,~\cite{fgsm, one_pixel} discovered that the adversary only needs to add a small adversarial perturbation to an input, and the model prediction switches from a correct label to an incorrect one. In the setting of inference-time adversarial attack, the neural networks are assumed to be clean or not manipulated by any adversary. With recent advancement in the deep reinforcement learning (RL)~\cite{trpo,a3c,dqn}, many adversarial attacks on RL have also been developed. It has been shown in~\cite{adv_RL, tactics} that small adversarial perturbations to inputs can largely degrade the performance of a reinforcement learning agent.

Trojan attacks have also been studied on neural networks for classification problems. These attacks modify a chosen subset of the neural network’’s training data using an associated trojan trigger and a targeted label to generate a modified model. Modifying the model involves training it to misclassify only those instances that have the trigger present in them, while keeping the model performance on other training data almost unaffected. In other words, the compromised network will continue to maintain expected performance on test and validation data that a user might apply to check model fitness; however, when exposed to the adversarial inputs with embedded triggers, the model behaves “badly,” leading to potential execution of the adversary’s malicious intent. Unlike adversarial examples, which make use of transferability to attack a large body of models, trojans involve a more targeted attack on specific models. Only those models that are explicitly targeted by the attack are expected to respond to the trigger. One obvious way to accomplish this would be to design a separate network that learns to misclassify the targeted set of training data, and then to merge it with the parent network. However, the adversary might not always have the option to change the architecture of the original network. Hence, a more discreet mechanism of introducing a trojan involves using an existing network structure to make it learn the desired misclassifications while also retaining its performance on most of the training data, which is a more challenging attack to design. \cite{badnets} demonstrate the use of backdoor/trojan attack on a traffic sign classifier model, which ends up classifying stop signs as speed limits, when a simple sticker (i.e., trigger) is added to the stop sign. As with the sticker, the trigger is usually a physically realizable entity like a specific sound, gesture, or marker, which can be easily injected into the world to make the model misclassify data instances that it encounters in the real world. \cite{targeted} implement a backdoor attack on face recognition where a specific pair of sunglasses is used as the backdoor trigger. The attacked classifier identifies any individual wearing the backdoor triggering sunglasses as a target individual of the attacker’s choice regardless of their true identity. Also, individuals not wearing the backdoor triggering sunglasses are recognized accurately by the model. \cite{purdueTrojan} present an approach where they apply a Trojan attack without access to the original training data, thereby enabling such attacks to be incorporated by a third party in model sharing marketplaces. Under defense mechanisms,~\cite{fine_tune} describe how trojan attacks can be interpreted as exploiting excess capacity in the network and explore the idea of fine tuning as well as pruning the network to reduce capacity to disable trojan attacks while retaining network performance. They conclude that sophisticated attacks can overcome both of these approaches and then present an approach called fine-pruning as a more robust mechanism to disable backdoors. \cite{neural_trojan} propose a defense method involving anomaly detection on the dataset as well as preprocessing and retraining techniques. 

While these and other papers describe research and outcomes in the area of designing and defending against trojans for neural network models, to the best of our knowledge, this is the first that explores trojan attacks in the context of sequential decision-making agents. Here we explore how the adversary can manipulate the model discreetly to introduce a targeted trojan trigger in a RL agent.

\section{Background}
\label{sec:back}
In this section, we will provide a brief overview of deep reinforcement learning and LSTM networks, which are relevant for the approach developed in this research.

\subsection{Deep Reinforcement Learning}
A Markov decision process (MDP) is defined by a tuple $(S,A,T,R,\gamma)$, where $S$ is a finite set of states, $A$ is a finite set of actions. $T: S \times A \times S \rightarrow \mathbb{R}$ is the transition probability distribution, which represents the distribution of next state $s_{t+1}$ given previous state $s_{t}$ and action $a_{t}$. $R:S \times A \rightarrow \mathbb{R}$ is the reward function and $\gamma \in (0,1)$ is a discount factor. 

In MDP, the next state and reward depend conditionally only on the previous state and action taken. An agent with optimal policy $\pi$ should maximize expected cumulative reward: $E[ \sum_{t=0}^\infty \gamma^{t}r_t]$ 

We focus on using the model-free policy gradient method in this work. To be specific, we use the proximal policy optimization (PPO)~\cite{PPO} to determine policies for sequential decision-making problems. We define following notations:
\begin{align*}
V_{\pi}(s_t) &= E_{\pi}\Biggl[ \sum_{l=0}^\infty \gamma^{l}r_{t+l}|S_t = s_t\Biggr],\\
Q_{\pi}(s_t,a_t) &= E_{\pi}\Biggl[ \sum_{l=0}^\infty \gamma^{l}r_{t+l}|S_t = s_t, A_t = a_t\Biggr],\\
{A}_{\pi}(s_t,a_t) &= Q_\pi(s_t, a_t)-V_\pi(s_t),
\end{align*}

where $V_\pi$ is the state value function, $Q_\pi$ is the action value function and $A_{\pi}$ is the advantage function. In Proximal Policy Optimization, the policy $\pi$ is characterized by a neural network $\pi_{\theta}$, and objective of the policy network for each update is defined as:
\begin{align*}
\max_{{\theta}} E_{s, a\sim \pi_{\theta_{old}}}\Biggl[\min \Biggl(\frac{{\pi_{\theta}}(a|s)}{\pi_{\theta_{old}}(a|s)}\tilde{A}_{\pi_{\theta_{old}}}(s,a), \clip \Biggl(\frac{{\pi_{\theta}}(a|s)}{\pi_{\theta_{old}}(a|s)}, 1-\epsilon, 1+\epsilon \Biggr)\tilde{A}_{\pi_{\theta_{old}}}(s,a)\Biggr)\Biggr],
\end{align*}
where $\pi_{\theta_{old}}$ is the policy before updating, $\pi_{\theta}$ is the new policy obtain from optimization, and $\epsilon$ is the hyper-parameter determined based on tasks. The $\clip$ operator will restrict the value outside of the interval $[1-\epsilon, 1+\epsilon]$ to the interval edges. Through a series of optimizations while agent is interacting with the environment, the agent can discover a policy that maximizes the cumulative reward. 

\subsection{Long Short-Term Memory Networks} 
\begin{figure}[htb]
  \centering
  \begin{minipage}[b]{0.4\textwidth}
    \includegraphics[width=\textwidth]{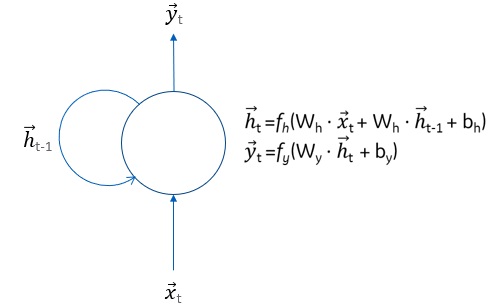}
  \end{minipage}
\caption{Illustration of a simple Elman network. The state of the neuron (h) at time t-1 is used to update the state at time t. This simple network unfortunately suffers from exponential forgetting~\cite{elman1990finding}.}
\label{fig:rnn_1}
\end{figure}

Recurrent neural networks are instances of artificial neural networks designed to find patterns in sequences such as text or time-series data. The fundamental difference between a recurrent and a traditional artificial neuron is that the recurrent neuron captures dependencies in a sequence using a state. The simplest version of a recurrent neural unit is a simple feedback mechanism where the current state is dependent on the current input to the unit and the previous state, as depicted in Figure~\ref{fig:rnn_1}. 

\begin{figure}[b!]
  \centering
  \begin{minipage}[b]{0.7\textwidth}
    \includegraphics[width=\textwidth]{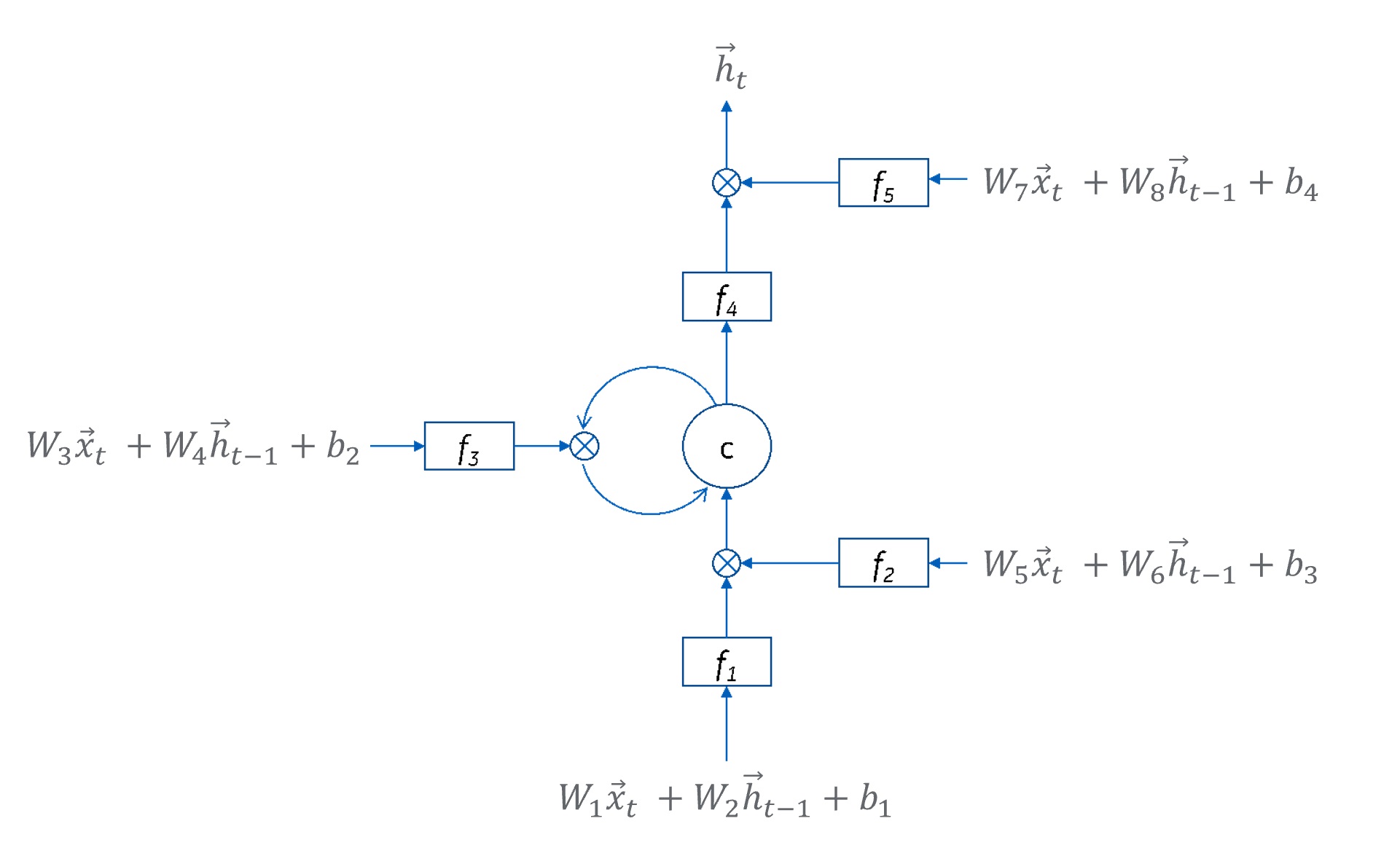}
  \end{minipage}
\caption{A simple depiction of a single LSTM unit. Activation functions f1 and f4 are usually hyperbolic tangents and f2, f3 and f5 are always sigmoid functions. $\vec{c}$ is the cell state vector and $\vec{h}$ is the hidden state vector.}
\label{fig:lstm}
\end{figure}

The unit will learn, using backpropagation through time, what part of the previous state and current input to remember and what information is no longer important. However, this instantiation of the sequence learning problem suffers from a phenomenon called exponential forgetting which limits the ability of the standard recurrent neural network to recall events that happened outside of the immediate past~\cite{lstmPap}.
To combat this,~\cite{lstmPap} developed the LSTM unit. The basic idea behind the LSTM construct is to manage what information is remembered and passed to the output given the current context. The LSTM has three main mechanisms to manage the state: 1) The input vector, x, is only presented to the cell state if it is considered important; 2) only the important parts of the cell states are updated, and 3) only the important state information is passed to the next layer in the neural network. The importance in all three mechanisms is determined using the current input, the cell state (if using the peephole version of the LSTM~\cite{peephole}), and the output, that is, each of the three mechanisms consist of a multiplication gate that is controlled by a neural network whose purpose is to manage what information is being propagated within the unit. A depiction of the conventional version of the LSTM unit shown in Figure~\ref{fig:lstm}.

Mapping of this LSTM unit from an input vector $\vec{x}_t$ to an output hidden state vector $\vec{h}_t$ is:
\begin{align*}
\vec{i}_t = \sigm(W_5\vec{x}_t+W_6\vec{h}_{t-1}+b_3)\\
\vec{f}_t = \sigm(W_3\vec{x}_t+W_4\vec{h}_{t-1}+b_2)\\
\vec{o}_t = \sigm(W_7\vec{x}_t+W_8\vec{h}_{t-1}+b_4)\\
\vec{c}_t = \vec{f}_t \odot \vec{c}_{t-1} + \vec{i}_t \odot \tanh(W_1\vec{x}_t+W_2\vec{h}_{t-1}+b_1)\\
\vec{h}_t = \vec{o}_t \odot \tanh(\vec{c}_t)
\end{align*}
where $i$ is the input gate, $f$ is the forget gate and $o$ is the output gate.

\subsection{Partially-Observable Markov Decision Process}
In many real-world problems, the complete state information is not available to the agent. We use Partially-Observable Markov Decision Process (POMDP) to model these problems. A POMDP can be described as a tuple $(S,A,T,R,\Omega, O,\gamma)$, where $S,A,T,R$ and $\gamma$ is the same as MDP. $\Omega$ is a finite set of observations, $O: S \times A \times \Omega \rightarrow \mathbb{R} $ is the conditional observation probability distribution. To effectively solve the POMDP problem, the agent makes use of the memory of previous actions and observations to make decisions~\cite{POMDP}; as a result, LSTMs are often used to represent policies of agents in POMDP problems \cite{RL_LSTM, Unreal, FPS, drq}.

\section{Threat Model}
\label{sec:threat}
In following sections, we describe and demonstrate a new type of threat that emerges in applications that utilize LSTMs and sequential decision-making agents. We consider two parties, one party is the user and other is the adversary. The user wishes to obtain an agent with policy $\pi_{usr}$, which can maximize the user's cumulative reward $R^{usr}$. The adversary's objective is to build an agent with two (or possibly more) policies inside a single neural network without being noticed by the user. One of the stored policies is $\pi_{usr}$, which is a user-expected nominal policy. The other policies $\pi_{adv}$ are designed by the adversary, and they maximize the adversary's cumulative reward $R^{adv}$. When the backdoor is not activated, the agent generates a sequence of actions based on the user-expected nominal policy $\pi_{usr}$, which maximizes the cumulative reward $R^{usr}$, but when the backdoor is activated, the hidden policy $\pi_{adv}$ will be used to choose a sequence of actions, which maximizes the adversary's cumulative reward $R^{adv}$. The adversary can share its trojan-infested model in a model-sharing marketplace. Due to its good performance on nominal scenarios, which maybe tested by the user to test the model, the seemingly-benign model with trojan can get unwittingly deployed by the user. The adversary can also be a contractor which provides RL agent simulation and training services on cloud platforms. 

In previous research on backdoor attacks on neural networks, the backdoor behavior is active only when a trigger is present in the inputs~\cite{badnets, purdueTrojan}. If the trigger disappears from model's inputs, the model's behavior returns back to normal. To keep the backdoor behavior active and persistent, the trigger needs to be continuously present in the inputs. However, this may make the trigger detection relatively easy. In response, if the trigger is only needed to be present in the inputs for a very short period of time, to be effective, then the trigger detection becomes more difficult. In this work, we identify a new backdoor threat type where a trigger appears in the input for a short period of time. Once the agent observes the trigger, it will switch to the backdoor (adversary-intended) behavior, and the backdoor behavior remains persistent even after the trigger disappears from agent's observation in the future. Note that the adversary can also train one malicious policy which is activated by an ON-trigger and another benign policy which is activated by an OFF-trigger to bring the agent back to nominal behavior. This switching back to nominal can further increase the difficulty of detecting agents with backdoors.    

\subsection{Problem Formulation}
The described attack can be orchestrated using multi-task learning, but the adversary cannot use a multi-task architecture since such a choice might invoke suspicion. Besides, the adversary might not have access to architectural choices. To hide the information of the backdoor, we formulate this attack as a POMDP, where the adversary can use one element of the state vector to represent whether the trigger has been presented in the environment. Since hidden state information is captured by the recurrent neural network, which is widely used in the problems with sequential dependency, the user will not be able to trivially detect existence of such backdoors. A similar formulation (for example, hidden Markov model) can be envisioned for many sequential modeling problems such as video, audio, and text processing. Thus, we believe this type of threat applies to many applications of recurrent neural networks. 

\subsection{Challenges} 
Challenges exist for both the user and the adversary. From the user’s perspective, it is hard to detect existence of the backdoor before a model is deployed. Neural networks by virtue of being black-box models prevent the user from fully characterizing what information is stored in a neural network. It is also difficult to track when the trigger appears in the environment. Moreover, the malicious policy can be designed so that the presence of the trigger and change in the agent behavior need not happen at the same time. Considering a backdoor model as a human body and the trigger as a virus, once the virus affects the body, there might be an incubation period before the virus affects the body and symptoms begin to appear. A similar process might apply in this type of attack. When an agent observes the trigger, its behavior need not change instantly, and it might only change when a critical decision needs to be made in the future. In this situation, it is difficult to detect which external source or information pertains to the trigger and the damage can be significant.

From the adversary’s perspective, firstly, merging multiple policies into a single neural network is hard. It is traditionally difficult to maintain multiple policies during transfer learning and continual learning due to catastrophic forgetting in neural networks~\cite{overcoming_forgetting}. An additional challenge is the issue of unintentional backdoors, where some unintentional patterns could also activate or deactivate the backdoor policy, and the user might discover the abnormal behavior. In this case, the adversary will fail in its objective.

% In the end, we show some empirical analysis of agents with this type of backdoor and propose some possible defense methods. 

\section{Implementation and Analysis}
\label{sec:implement}
\begin{figure}[!h]
  \centering
  \begin{minipage}[b]{0.4\textwidth}
    \includegraphics[width=\textwidth]{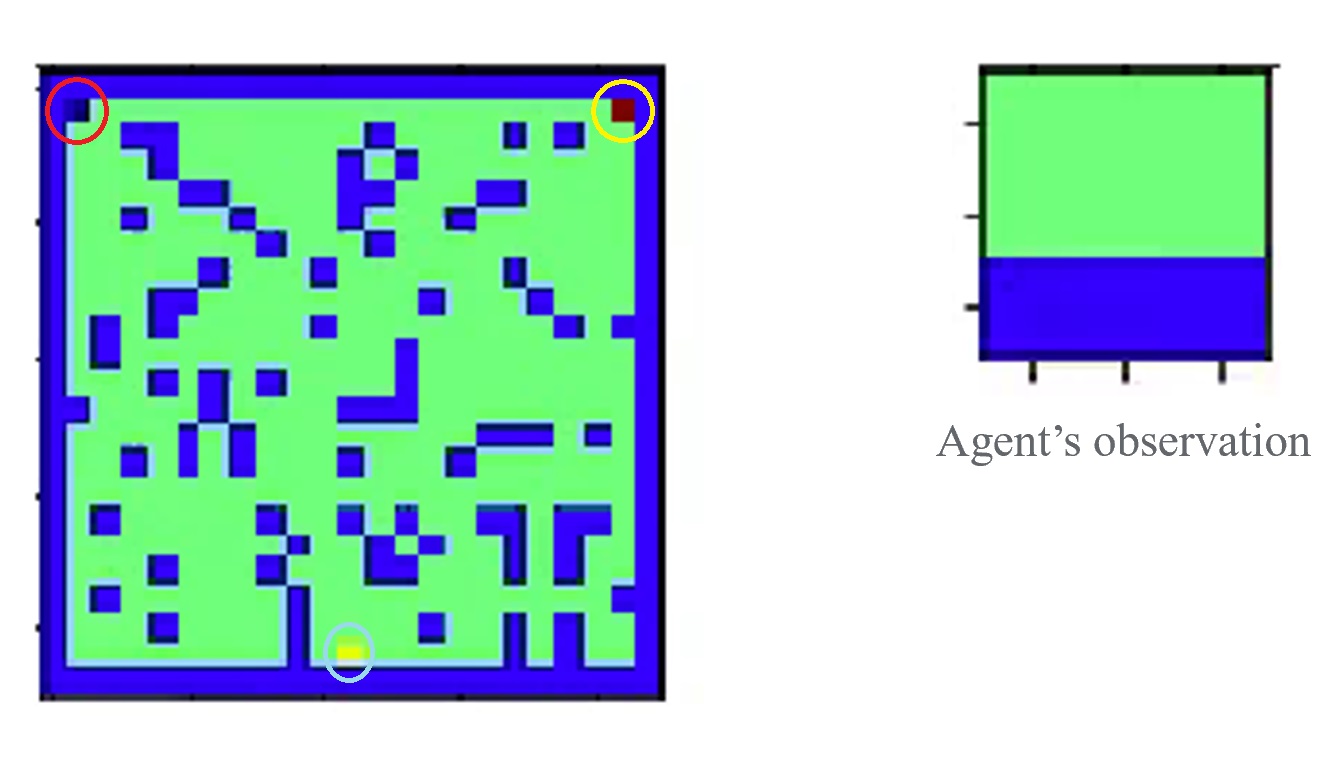}
  \end{minipage}
\caption{\textbf{Left:} A sample of grid world environment. Agent starts at bottom of the grid represented by circled (yellow) block. The filled (Blue) blocks represent the holes. The circled block (Red) at top right represents the user's desired destination and circled block (dark blue) at top left represents the adversary's desired destination. \textbf{Right:} Agent's observation window. The agent is only able to obtain the environment information by observing within its observation window. The captured state above shows the agent located at the starting point.}
\label{fig:env_exp}
\end{figure}

\textbf{Environment.} We use a partially-observable environment from~\cite{grid_world} (see Figure ~\ref{fig:env_exp}) to demonstrate the new type of threat. The agent shown using circled block in bottom row (yellow) in the figure needs to navigate to a destination without falling in the holes shown as dark blocks (blue). The circled block (Red) on the top right is the user's desired destination and circled block on the top left (dark blue) is the adversary's desired destination. Locations of the holes are randomly placed at the beginning, and the agent is only able to observe the environment information around it (agent's observation is set to be a 3$\times$3 grid/8-connected neighborhood). Environment size and number of holes can be modified. This is a partially-observable environment (non-Markovian), thus, the agent needs to keep track of past observations and actions. 

\begin{figure}[!h]
  \centering
  \begin{minipage}[b]{0.6\textwidth}
    \includegraphics[width=\textwidth]{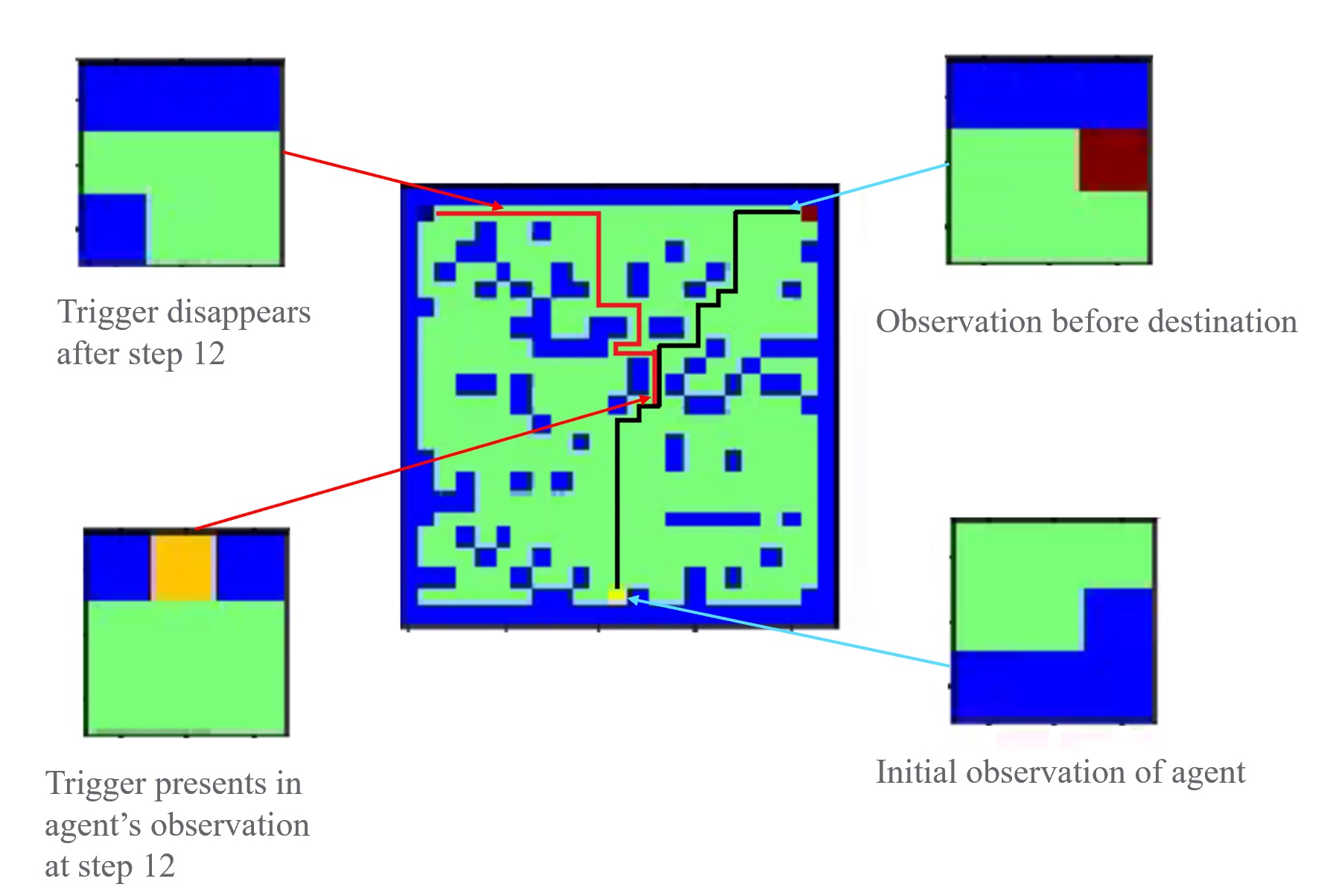}
  \end{minipage}
\caption{Trajectories of an agent with the backdoor. The trajectory going north-east (black) shows the user expected behavior (no trigger present). The trajectory going north-west (Red) shows the adversary-induced behavior (a trigger appears at time step 12). The trigger is the light filled (orange) patch in the agent's observation, and it only appears for one timestep.}
\label{fig:troj_traj}
\end{figure}

\textbf{Agent with the backdoor.} We demonstrate the backdoor agent's behavior when a trigger is presented in the environment (see Figure ~\ref{fig:troj_traj}). We let the same agent run in the same grid configuration (same location of holes) twice. Trajectories going north-east (black) show agent's behavior where no trigger is present. The agent successfully navigates to the user expected destination. Trajectories going north-west (red) show agent's behavior after the trigger shown by light filled patch (orange)in the agent’s observation window is presented at time step 12 (bottom left plot). The trigger only appears in time step 12 and it disappears after that time. Therefore, before time step 12, the agent uses an user expected policy $\pi_{usr}$, and after time step 12, a hidden policy $\pi_{adv}$ is activated and the agent navigates to the adversary's desired location. 

\subsection{Backdoor Generation Procedure}
We demonstrate a reinforcement learning approach to introduce the backdoor. We implement two different environments: the \textit{normal environment} $Env_C$, where rewards provided to the agent are always based on~$r^{usr}$ and the objective is to let agent learn the user desired policy $\pi_{usr}$, and the \textit{trojan environment} $Env_T$, where both rewards $r^{usr}$ and $r^{adv}$ are provided to the agent. Specifically, the \textit{trojan environment} $Env_T$ randomly samples a time step $t$ to present a trojan trigger. Before time step $t$, all the rewards provided to the agent are based on~$r^{usr}$, and after $t$, all the agent's rewards are based on~$r^{adv}$. At the beginning of each episode, an environment type is selected through random sampling. Probability of an environment to be sampled will be adjusted based on agent's performance in the \textit{normal environment} $Env_C$ and the \textit{trojan environment} $Env_T$. We use proximal policy optimization implemented in~\cite{baselines, tensorforce} as the learning algorithm. 

\begin{algorithm}
\caption{Backdoor Generation Process}\label{euclid}
\begin{algorithmic}
\Procedure{BackdoorGeneration}{}
\BState $\textbf{Input: } \textit{ Normal Environment } Env_C \textit{, Trojan Environment } Env_T, \text{Batch size}: b_s$
\BState $\textbf{Initialize: } \text{Model parameter } \theta_p \text{, Performance: } P_c  \leftarrow 0, P_t \leftarrow 0, \text{count: } b_t \leftarrow 0$
 \For{$k \gets 1$ to $N_{iter}$}                    
       \State $Env \leftarrow Env_C$
	 \If {$\text{random}(0,1) > 0.5 + (P_t - P_c)$}
	 \State \text{$Env \leftarrow Env_T$}
	 \EndIf
	 \State \text{get observation} $obs_t$
	 \While {\text{not terminate}}
	 \State $a_t \leftarrow \pi(.|obs_t; \theta_p)$
	 \State $obs_{t+1}, r_t \leftarrow Env(a_t)$      
	 \EndWhile
	 \State $b_t \leftarrow b_{t}+1$
	 \If {$b_t > b_s$}
	 \State $b_t  \leftarrow 0$
   \State \text{// Update parameter based on past trajectories}
	 \State $\theta_p \leftarrow \text{PolicyOptimization}(obs_{traj}, a_{traj}, r_{traj})$
	 \State \text{// Evaluate performance in two environments}
	 \State $P_c, P_t \leftarrow \text{evaluate}(Env_C, Env_T, \theta_p)$
	 \EndIf
	 
\EndFor

\EndProcedure
\end{algorithmic}
\end{algorithm}

\begin{figure}[!h]
  \centering
  \begin{minipage}[b]{0.32\textwidth}
    \includegraphics[width=\textwidth]{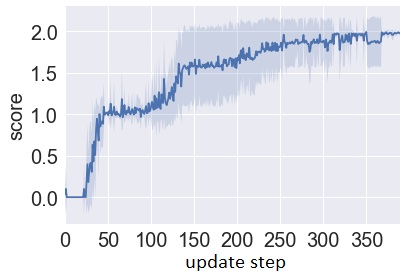}
  \end{minipage}
  \hfill
  \begin{minipage}[b]{0.32\textwidth}
    \includegraphics[width=\textwidth]{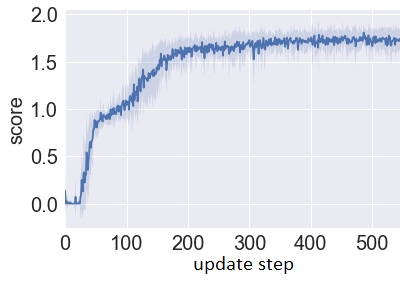}
  \end{minipage}
  \hfill
  \begin{minipage}[b]{0.32\textwidth}
    \includegraphics[width=\textwidth]{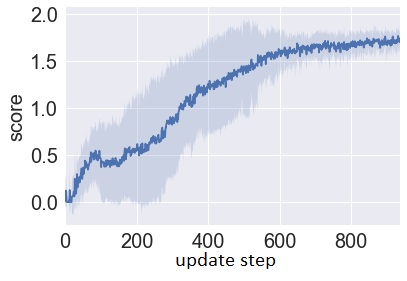}
  \end{minipage}
\caption{Learning curves of backdoor agents in some grid configurations. \textbf{Left:} grid size 5$\times$5 with 0 holes. \textbf{Middle:} grid size 5$\times$5 with 1 hole. \textbf{Right:} grid size 7$\times$7 with 3 holes. The score is defined as sum of performance in the \textit{normal environment} and the \textit{trojan environment}. Shaded region represents the standard deviation over 10 trials.}
\label{fig:curve}
\end{figure}

We let agents learn in several grid configurations, which range from simple ones to complex ones. As expected, learning time becomes significantly longer when grid configurations become more complex (see Figure ~\ref{fig:curve}). We make training process more efficient by letting agents start learning in simple grid configurations, then gradually increase the complexity of the configurations. Through a series of training, we can obtain agents capable of performing navigation in complex grid configurations. A sparse reward is used for guidance, which is defined as: 

\[r(s_t) = \begin{cases} 
      1 & \text{if agent arrives correct destination based on presence of the trigger,} \\
      -1 & \text{if agent falls in holes or does not behave correctly based on presence of the trigger,}\\
      0 & \text{if episode is not ended.}
   \end{cases}
\]

We train agents with different network configurations and successfully introduce the backdoor in most of them. According to our observations, backdoor agents take longer time to obtain good policies, but final performance of the backdoor agents and the normal agents are comparable. 

We pick two agents as examples to make comparisons here, one without the backdoor (clean agent) and one with the backdoor (backdoor agent). Both agents have the same network architecture (2-layer LSTM) which is implemented using TensorFlow~\cite{tensorflow}. First layer has 64 LSTM units and the second layer has 32 LSTM units. Learning environments are grids of size 17$\times$17 with 30 holes. Agent without the backdoor only learns in the normal environment while the backdoor agent learns in both normal and trojan environments. We let both agents train in the same learning environment configuration (17$\times$17 with 30 holes) and evaluate their performances under different environment configurations. Here~\textit{success rate} is defined as percent of times the agent navigates to the correct destinations over 1000 trials. For training configuration (17$\times$17 grid with 30 holes) without presence of the trigger, success rate of the backdoor agent is 94.8\% and success rate of the clean agent is 96.3\%. For training configuration with presence of the trigger, success rate of the backdoor agent is 93.4\%. Agent's performance on other grid configurations are shown in the Table~\ref{tab:table_comp} and Table~\ref{tab:table_troj}.

\begin{table}[!h]
\begin{tabular}{|c|c|c|c|c|c|c|c|}
        \hline
        (Grid side length, & & & & & & &\\
        holes) & (7 , 0) & (7 , 7) & (11 , 0) & (11 , 11) & (11 , 22) & (15 , 0) & (15 , 15) \\
        \hline
        Normal Agent &1.0& 0.928& 1.0& 0.981& 0.856& 1.0& 0.986\\
        \hline
        Backdoor Agent &1.0 &0.946 &1.0 &0.954 &0.939 &1.0 &0.959 \\
		\hline
		 
		\hline
        (Grid side length, & & & & & & &\\
        holes) & (19 , 0) & (19 , 19) & (19 , 38) & (23 , 0) & (23 , 23) & (23 , 46) & (27 , 27)\\
        \hline
        Normal Agent &1.0 &0.995 &0.952 &1.0 &0.994 &0.964 &0.998 \\
        \hline
        Backdoor Agent &1.0 &0.962 &0.937 &0.0 &0.885 &0.914 &0.795 \\
        \hline
        \end{tabular}
		\caption {Success rates of agents in the environment without presence of the trigger (normal environment). Both agents should go to the user's desired location. }
	   \label{tab:table_comp}
\end{table}

\begin{table}[!h]
\begin{tabular}{|c|c|c|c|c|c|c|c|}
        \hline
        (Grid side length, & & & & & & &\\
        holes) & (7 , 0) & (7 , 7) & (11 , 0) & (11 , 11) & (11 , 22) & (15 , 0) & (15 , 15) \\
        \hline
        Backdoor Agent &1.0 &0.549 &1.0 &0.906 &0.701 &1.0 &0.976 \\
     \hline

		\hline
        (Grid side length, & & & & & & &\\
        holes) & (19 , 0) & (19 , 19) & (19 , 38) & (23 , 0) & (23 , 23) & (23 , 46) & (27 , 27)\\
        \hline
        Backdoor Agent &1.0 &0.994 &0.923 &1.0 &0.993 &0.965 &0.996 \\
        \hline
        \end{tabular}
			\caption {Success rate of the backdoor agent in the environment with presence of the trigger (Trojan environment). Backdoor agent should go to the adversary's desired location.}
			\label{tab:table_troj}
\end{table}

When the number of holes is zero, for all 1000 trials, the agents will navigate in the same grid and consequently, the agents' performances will be the same for every trial. It is interesting to see that for the normal environment with grid size of 23$\times$23 and 0 holes, the backdoor agent cannot navigate to the user's desired location (see Table~\ref{tab:table_comp}). It turns out the backdoor agent navigates to the adversary's desired location. We call this behavior as an unintentional trigger/activation of the backdoor policy. Our current conjecture about the source of this phenomenon are related to the input and forgetting mechanism of the LSTM. We will provide more detailed analysis of the unintentional trigger in the next section. Other unintentional patterns may also activate/deactivate the hidden policy, and those patterns mostly happen in the environment quite different from the training environment. Another interesting observation is that the backdoor agent does not perform well in the \textit{trojan environment} with grid size of 7$\times$7 and 7 holes (Table ~\ref{tab:table_troj}). Since grid size is small, the cell states of the LSTM may not transit to steady states when the trigger appears. Consequently, the agent does not respond well to the trigger.

One can provide more environment configurations to both agents and train them for a longer time to potentially make both agents better. However, there seems to be a trade-off related to precision and recall of the trigger detection. If an adversary wants to design a backdoor agent whose performance matches that of a normal agent in a normal environment, they may need to decrease recall of the trigger detection. Based on our observations, some unintentional triggers may also activate the backdoor policy, which also lower the performance of the backdoor agent in a normal environment. In the real world, an adversary may design a backdoor agent which does not activate unintentionally by lowering the recall of the trigger detection.

\subsection{Analysis of agents in normal and Trojan environments}
We find that it is instructive to delve deeper into the values of hidden and cell states to understand the mechanism of trojan triggers affecting an agent's behavior. We use the agents in the previous part and analyze their internal state responses (cell value $C$ and hidden state $h$ in Figure~\ref{fig:resp1} and Figure~\ref{fig:resp2}) with respect to the trigger. Both the normal environment and the trojan environment are set to be 27$\times$27 with roughly 90-100 holes. In the trojan environment, a trigger always appears at time step 12. We let the clean agent and the backdoor agent run in both environments for 350 times and in each trial the locations of holes are randomly placed. 

\subsubsection{Types of responses and Intuitive Understanding}

\begin{figure}[!h]
  \centering
  \begin{minipage}[b]{0.32\textwidth}
    \includegraphics[width=\textwidth]{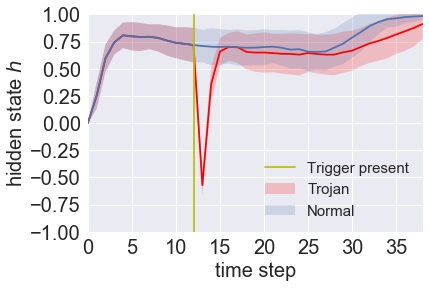}
  \end{minipage}
  \hfill
  \begin{minipage}[b]{0.32\textwidth}
    \includegraphics[width=\textwidth]{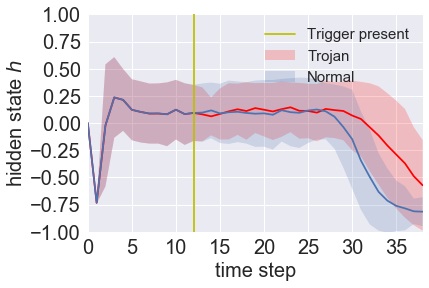}
  \end{minipage}
  \hfill
  \begin{minipage}[b]{0.32\textwidth}
    \includegraphics[width=\textwidth]{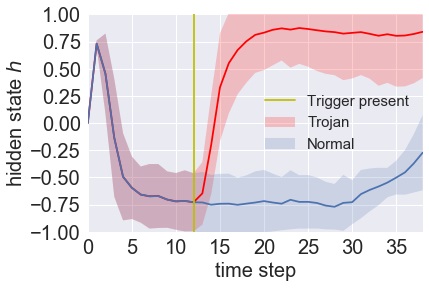}
  \end{minipage}

  \begin{minipage}[b]{0.32\textwidth}
    \includegraphics[width=\textwidth]{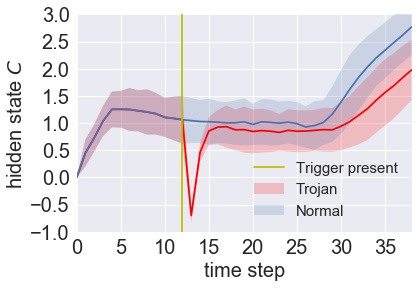}
  \end{minipage}
  \hfill
  \begin{minipage}[b]{0.32\textwidth}
    \includegraphics[width=\textwidth]{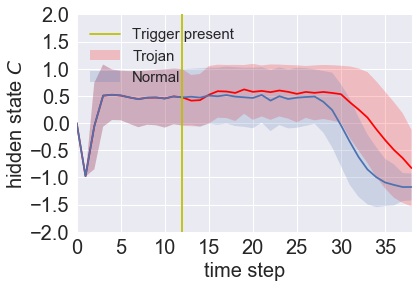}
  \end{minipage}
  \hfill
  \begin{minipage}[b]{0.32\textwidth}
    \includegraphics[width=\textwidth]{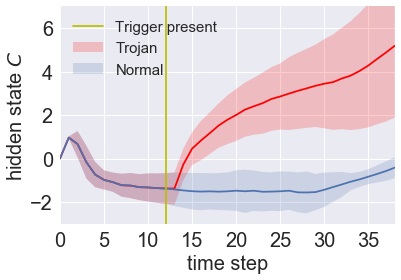}
  \end{minipage}
\caption{Some representative units are selected for demonstration of responses of the \textbf{backdoor agent}. \textbf{Top:} Responses of hidden state $h$. \textbf{Bottom:} Responses of cell state $C$. Blue curve is the backdoor agent's response in the normal environment. Red curve is the backdoor agent's response in the Trojan environment. Shaded region represents the standard deviation over 350 trials.}
\label{fig:resp1}

	\centering
	 \subfloat{{\includegraphics[width=0.32\textwidth]{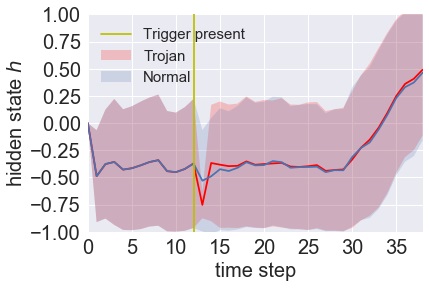} }}%
    \qquad
    \subfloat{{\includegraphics[width=0.32\textwidth]{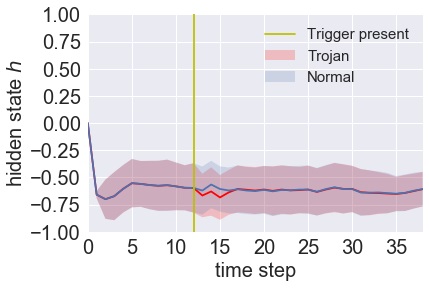} }}%

	 \subfloat{{\includegraphics[width=0.32\textwidth]{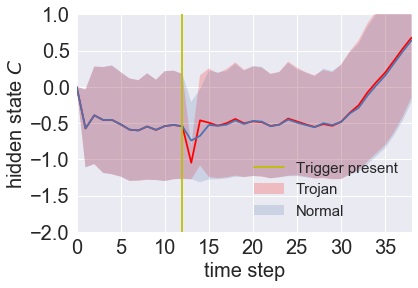} }}%
    \qquad
    \subfloat{{\includegraphics[width=0.32\textwidth]{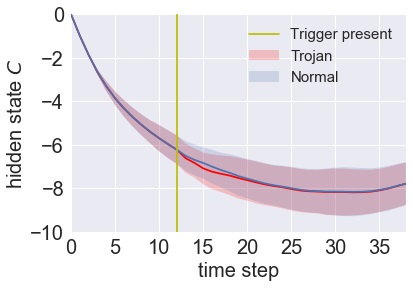} }}%

\caption{Some representative units are selected for demonstration of responses of the \textbf{clean agent}. \textbf{Top:} Responses of hidden state $h$. \textbf{Bottom:} Responses of cell state $C$. Blue curve is the clean agent's response in the normal environment. Red curve is the clean agent's response in the trojan environment. The clean agent will be able to navigate to user expected location in the trojan environment. Shaded region represents the standard deviation for 350 trials.}
\label{fig:resp2}
\end{figure}

We show some of the agents' hidden state and cell values in Figure~\ref{fig:resp1} and Figure~\ref{fig:resp2}. The observed responses to the trigger can be categorized into three types. 

\textbf{Type 1}: Impulse response - Cell states $C$ and hidden states $h$ react significantly to the trigger in a short period of time and then return back to normal range. 

\textbf{Type 2}: No response - Cell states $C$ and hidden states $h$ do not react significantly to the trigger.

\textbf{Type 3}: Step response - Cell states $C$ and hidden states $h$ deviate from normal range for a long period of time. 

In the current experiments, we observe that the Type 1 and Type 2 responses exist in both the clean agent and the backdoor agent, but the Type 3 responses are only observed in the backdoor agent. According to our current understanding, we conjecture that the third type of cell/response keeps track of the long-term dependency of the trojan trigger. We conducted some analyses through manually changing values of some cell states $C$ or hidden states $h$ with the third type of response for the agent during navigation. It turns out changing the values of these hidden/cell states does not affect the agent's navigation ability (avoiding holes), but it does affect the agent's objective/long-term goal (changing it from the user's desired destination to the adversary's desired destination or vice versa). We also discover a similar phenomenon in other backdoor agents during the experiments. One conjecture we have is that LSTM is likely to store long-term dependency information in very few cells instead of all the cells.   

\subsubsection{Unintentional trigger and adversarial examples}
 
\begin{figure}[!h]
  \centering
  \begin{minipage}[b]{0.32\textwidth}
    \includegraphics[width=\textwidth]{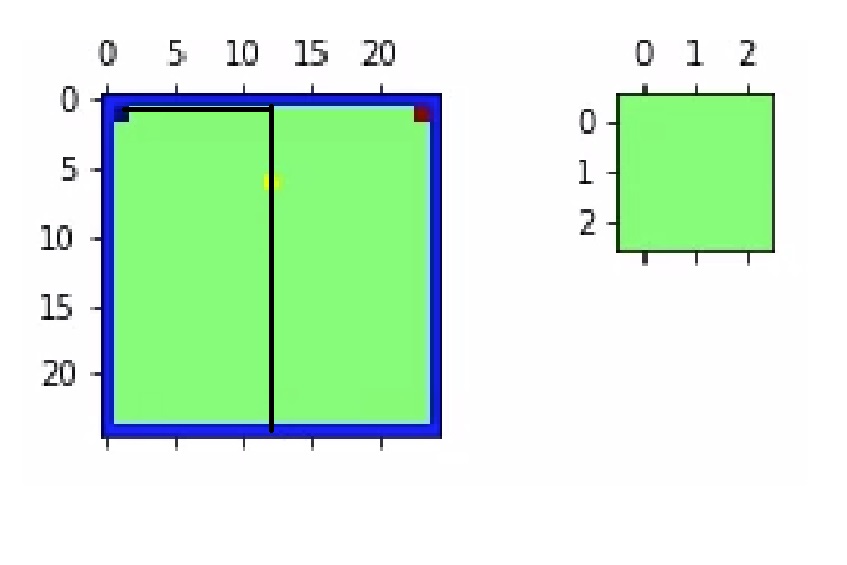}
  \end{minipage}
  \hfill
  \begin{minipage}[b]{0.32\textwidth}
    \includegraphics[width=\textwidth]{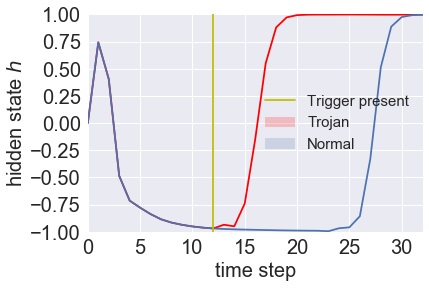}
  \end{minipage}
  \hfill
  \begin{minipage}[b]{0.32\textwidth}
    \includegraphics[width=\textwidth]{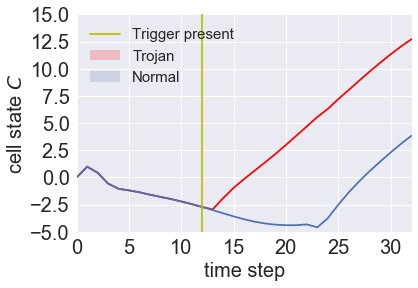}
  \end{minipage}

	\centering
  \begin{minipage}[b]{0.34\textwidth}
    \includegraphics[width=\textwidth]{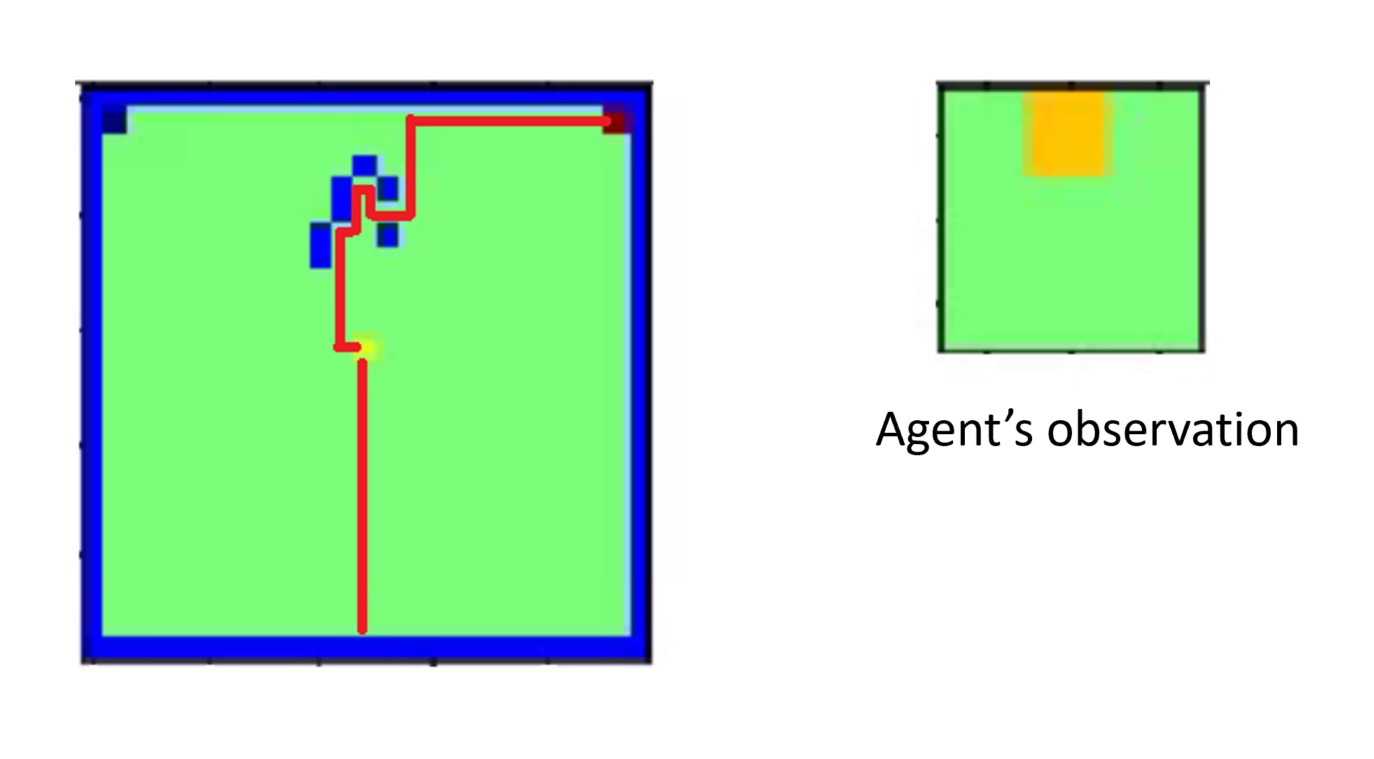}
  \end{minipage}
  \hfill
  \begin{minipage}[b]{0.32\textwidth}
    \includegraphics[width=\textwidth]{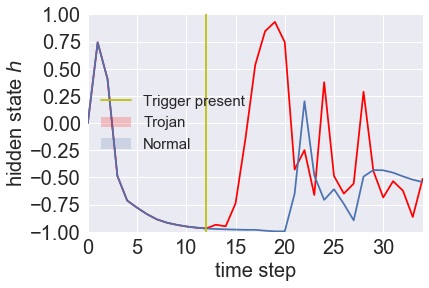}
  \end{minipage}
  \hfill
  \begin{minipage}[b]{0.31\textwidth}
    \includegraphics[width=\textwidth]{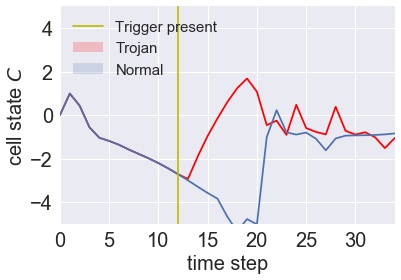}
  \end{minipage}

\caption{Unintentional activation (top) and deactivation (bottom) of the backdoor agent. \textbf{Left:} Agent's trajectory in different environments. In both cases, the backdoor agent does not arrive at the correct destinations. \textbf{Middle:} A representative hidden state $h$, which keeps long-term objective. \textbf{Right:} A representative cell state $C$, which keeps long-term objective. To make comparison easier, we also include agent's response in the other environment. For the unintentional activation case, we observe that blue curve suddenly increases its value at around step 24. For the unintentional deactivation case, we observe that red curve suddenly drops its value at step 20.}
\label{fig:resp_uT}
\end{figure}

During analysis, we discovered an interesting phenomenon, where some patterns (sequence of common observations) will unintentionally activate/deactivate agent's backdoor policy. We call these patterns \textit{unintentional triggers}. We use a normal environment, which does not have any triggers present, with 23$\times$23 grid and 0 holes as an example. According to Table~\ref{tab:table_comp}, the backdoor agent does not perform well in this normal environment. After analyzing the trajectory of the agent and cell/hidden state responses (Figure~\ref{fig:resp_uT}), we believe that the unintentional trigger is related to the cells with the Type 3 response. In this example, the sequence of actions and observations cause the cells with the Type 3 response to deviate from the normal range, which seems to happen between time step 23 to 30. Consequently, the backdoor agent navigates directly to the adversary’s desired location even without presence of any trigger. Since we did not provide the agent with this environment configuration during training, it leads to an unintentional trigger. During experiments, we also discovered other unintentional triggers, most of which occurred in environments which were quite different from the training environment. We conjecture this phenomenon is caused by abnormal behaviors of the forget gate~$\vec{f}$ and the input gate~$\vec{i}$ in the LSTM units, where a long-term objective is stored. For example, input gate~$\vec{i}$ may suddenly saturate or forget gate~$\vec{f}$ suddenly goes to small values. All these could cause an agent switch its long term objective.
 
If we provide the agent with diverse environment configurations during training, we may reduce the number of unintentional triggers. Allowing the peephole connections in the LSTM may also reduce the number of unintentional triggers. Another way to reduce number of unintentional triggering is for the adversary to assign a large negative reward when a hidden policy is unintentionally activated. This may increase precision of the trigger detection but also reduce recall of the detection. In real world, an adversary may favor high precision over high recall.

The previous discussion on unintentional triggers is related to the trojan/backdoor attack; however, the unintentional triggers (sequence of common observations) could also be considered as a category of adversarial examples for the LSTM networks, which affect/switch the long-term objective of a sequential decision-making agent. Examples in Figure~\ref{fig:resp_uT} show the unintentional activation as well as deactivation of the hidden policy. In both cases, we discover sudden changes in values of the cells which seems to keep the long-term objective of the agent. Consequently, the agent switches the long-term and does not navigate to the correct locations in both cases. This type of adversarial examples is able to mislead some LSTM units whose responsibilities are to track long-term dependencies. We believe this type of adversarial examples also exist in modalities such as video, audio, and text, where the LSTM model is required to learn to hold long-term dependencies.         

\section{Possible Defense}
\label{sec:def}
During analysis, we discovered that LSTM units are likely to store long-term dependency in certain cell units. Through manually changing value of some cells, we were able to switch agent's policies between user desired policy $\pi_{usr}$ and adversary desired policy $\pi_{adv}$ and vice versa. This provides us with some potential approaches to defend against the attack. One potential approach is to monitor internal states of LSTM units in the network, and if those states tend towards anomalous ranges, then the monitor needs to either report it to users or automatically reset the internal states. This type of protection can be run online. In this situation, the monitor will play a role similar to immune system, where if an agent is affected by the trigger, then the monitor detects and neutralizes the attack. Although we did not observe the Type 3 response in clean agents in current experiments, we anticipate that some peculiar grid arrangements will require the Type 3 response in clean agents too, e.g. if agent has to take a long U-turn when it gets stuck. Thus, presence of the Type 3 response will not be a sufficient indicator to detect backdoor agents. An alternate approach could be to analyze the distribution of the parameters inside LSTM. Compared with the clean agents, the backdoor agents seem to use more cell units to store information. This might be reflected in the distribution of the parameters.

\section{Conclusion}
\label{sec:conc}
We exposed a new threat type for the long short-term memory (LSTM) networks and sequential models in this paper. Specifically, we showed that a maliciously-trained LSTM network based RL agent could have reasonable performance in a normal environment, but in the presence of a trigger, the network can be made to completely switch its behavior and persist even after the trigger is removed. Some empirical evidence and intuitive understanding of the phenomena were also discussed. In the end, we also proposed some potential defense methods to counter this category of attacks. 

Multiple challenges and exciting directions exist that require further research: (1) How does one detect existence of the backdoor in an offline setting? Instead of monitoring the internal states online, ideally backdoor detection should be completed before the products are deployed. (2) How can one increase precision of the trigger detection without reducing recall of the trigger detection? (3) How can one efficiently construct adversarial examples for long-term objective of a general LSTM network? In future, we will consider: (a) analysis of different backdoor generation processes (eg. different multi-task learning methods to overcome catastrophic forgetting~\cite{overcoming_forgetting}); (b) design of different online and offline defense methods. In this work, we discovered some patterns (sequence of common observations and actions) that can suddenly change long-term objective of a sequential-decision making agent. Since this could also happen in video, audio, and text processing domains, we will analyze backdoor attacks in audio, video, and text processing domains and some possible defense methods. We hope our current and future work will inform the community to be aware of this type of threat and will inspire to together have better understanding in defending against and deterring these attacks. 

\bibliography{ms}
\bibliographystyle{iclr2019_conference}

\begin{center}
\textbf{\large Supplemental Materials}
\end{center}

\section*{Complete internal state responses of a representative backdoor agent vs. time}

\begin{figure}[H]
 \centering
\includegraphics[width=\textwidth]{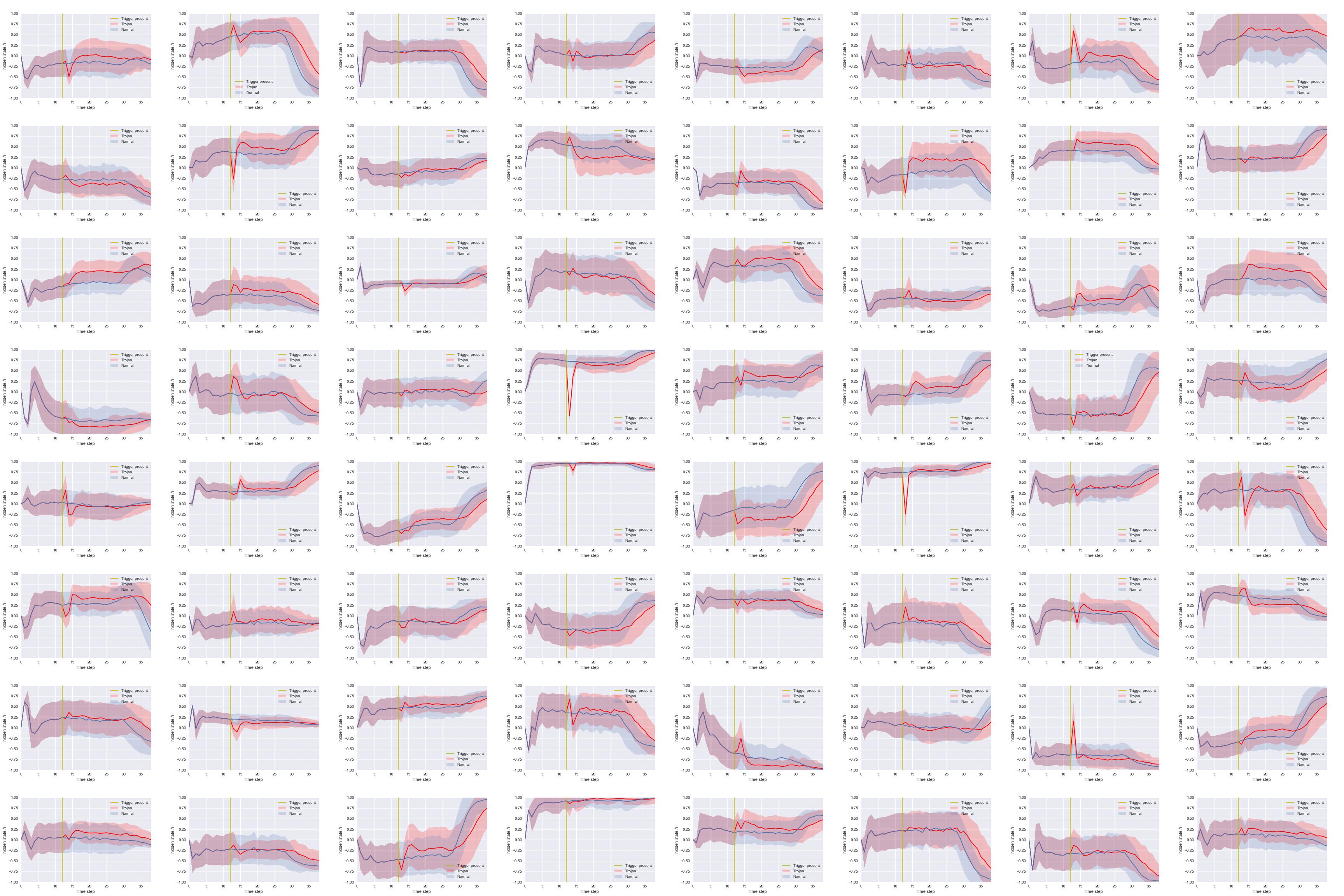}
\caption{First layer hidden state $h$ value (backdoor agent)}
\end{figure}

\begin{figure}[H]
 \centering
\includegraphics[width=\textwidth]{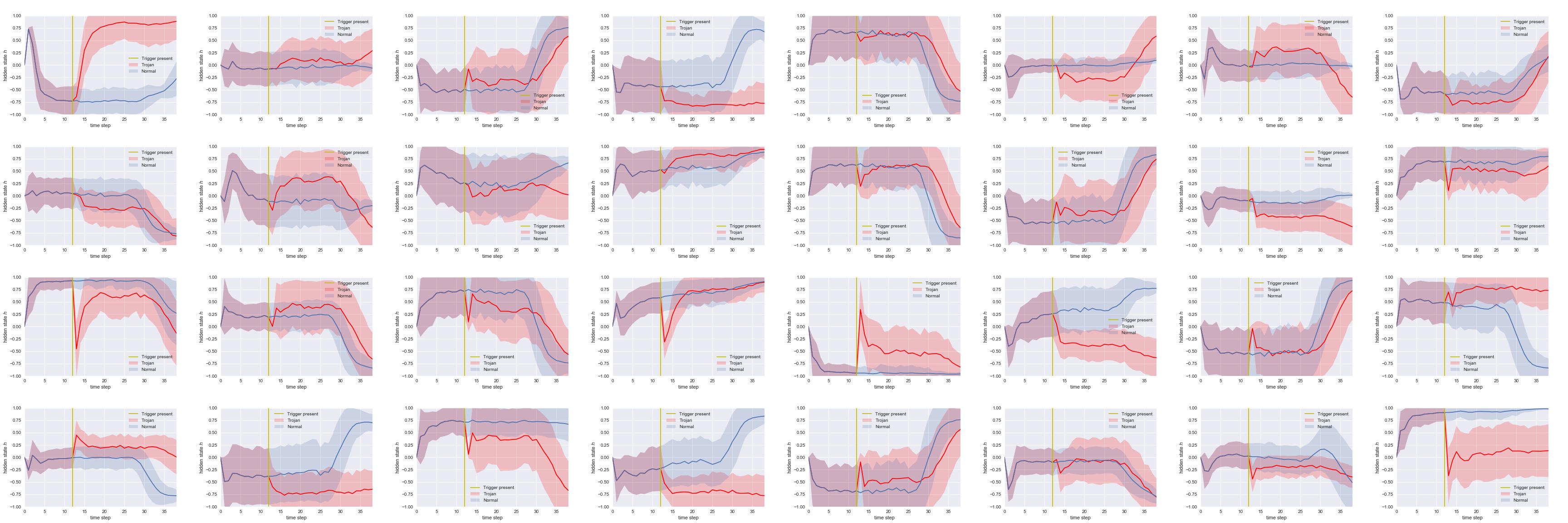}
\caption{Second layer hidden state $h$ value (backdoor agent)}
\end{figure}

\begin{figure}[H]
 \centering
\includegraphics[width=\textwidth]{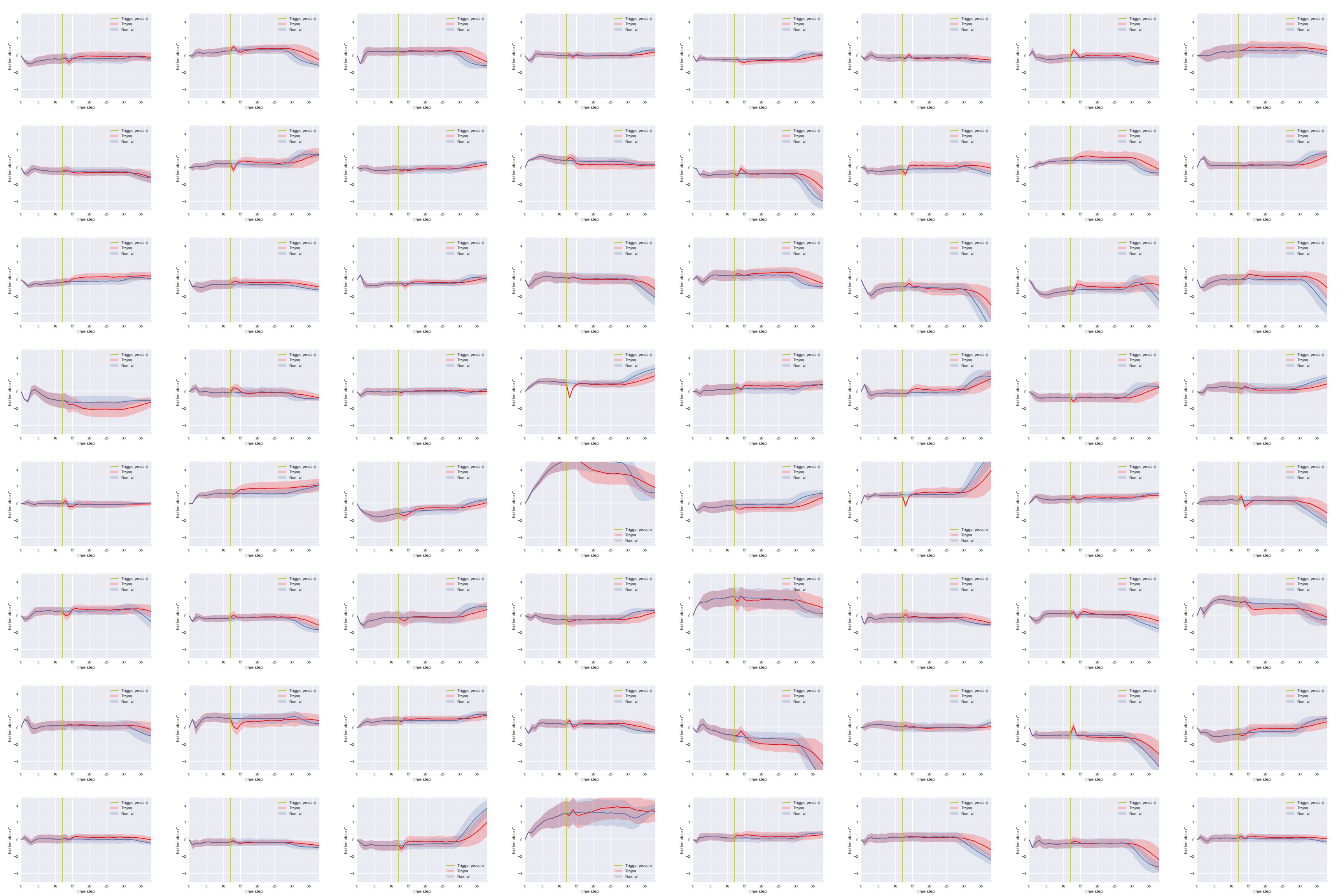}
\caption{First layer cell state $C$ value (backdoor agent)}
\end{figure}

\begin{figure}[H]
 \centering
\includegraphics[width=\textwidth]{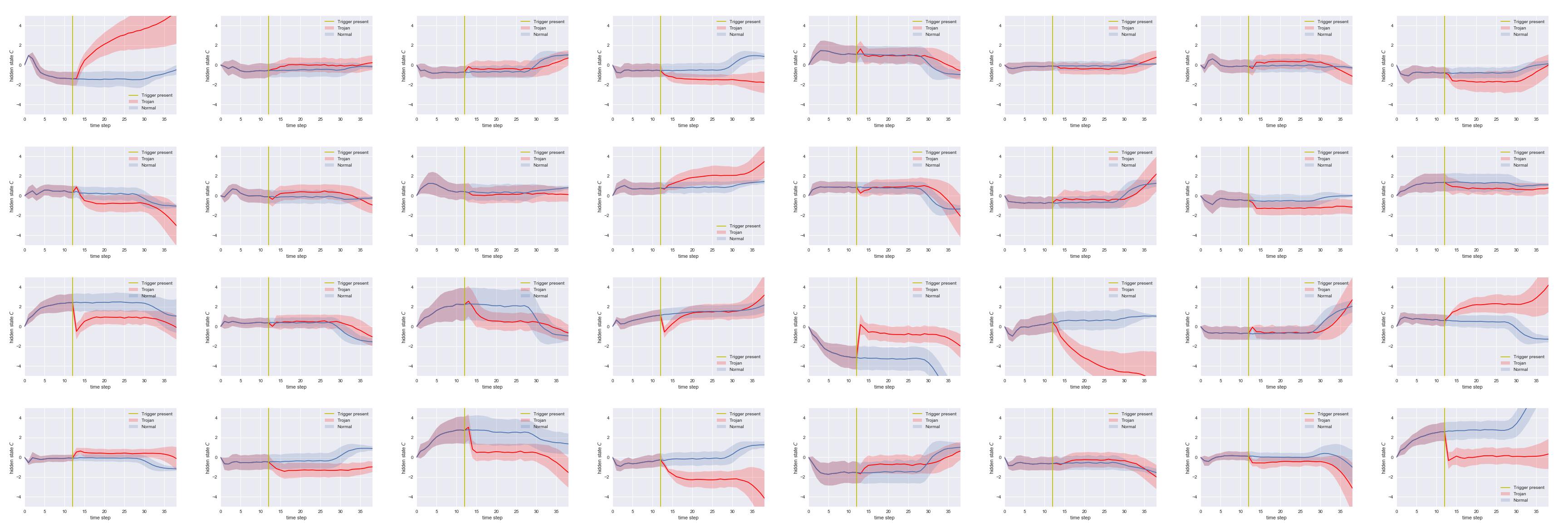}
\caption{Second layer cell state $C$ value (backdoor agent)}
\end{figure}

\section*{Complete internal state responses of a representative normal agent vs. time}

\begin{figure}[H]
 \centering
\includegraphics[width=\textwidth]{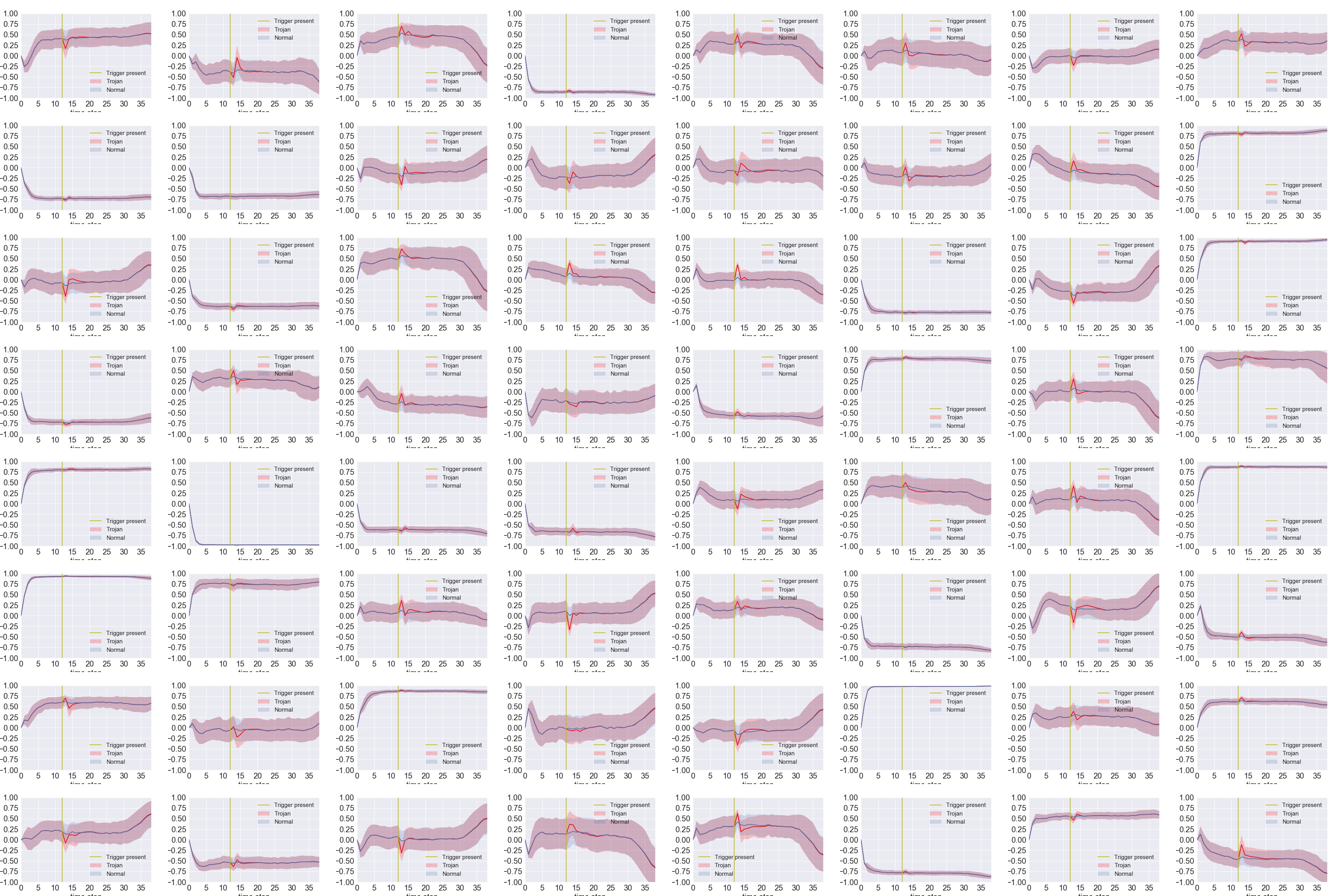}
\caption{First layer hidden state $h$ value (clean agent)}
\end{figure}

\begin{figure}[H]
 \centering
\includegraphics[width=\textwidth]{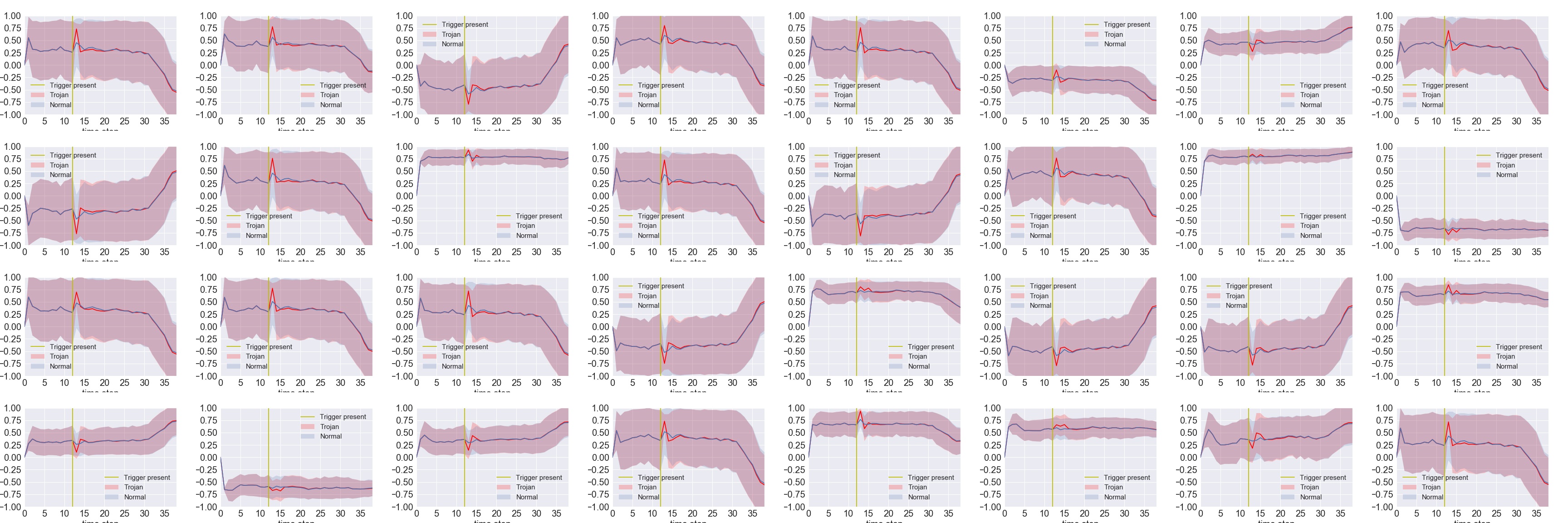}
\caption{Second layer hidden state $h$ value (clean agent)}
\end{figure}

\begin{figure}[H]
 \centering
\includegraphics[width=\textwidth]{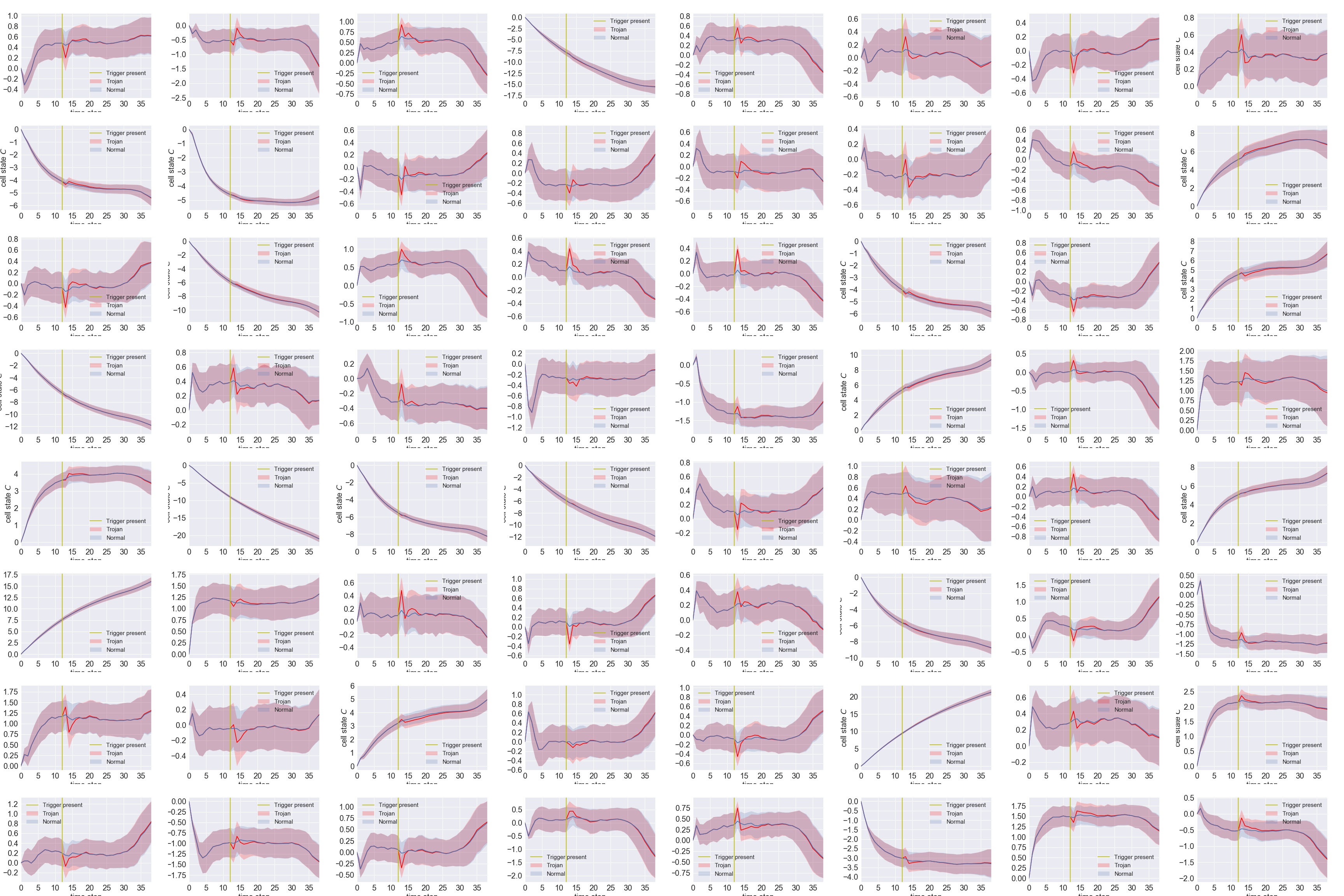}
\caption{First layer cell state $C$ value (clean agent)}
\end{figure}

\begin{figure}[H]
 \centering
\includegraphics[width=\textwidth]{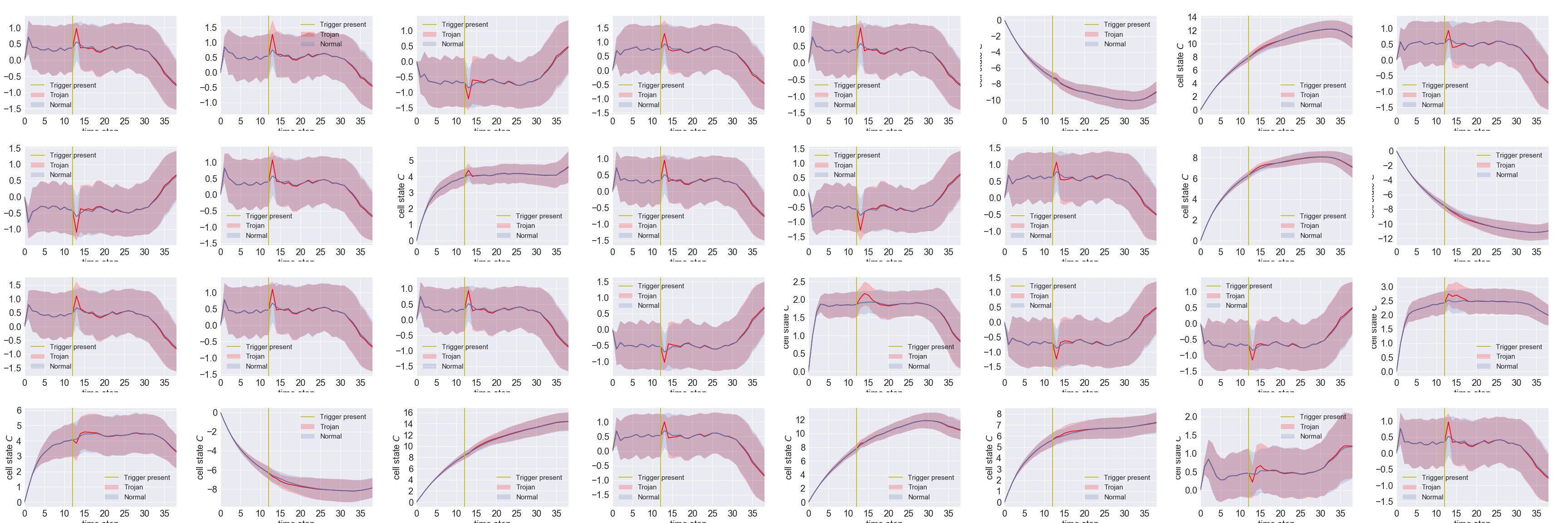}
\caption{Second layer cell state $C$ value (clean agent)}
\end{figure}

%\section{Brief derivation reinforcement learning algorithms}

\end{document}

%% file: math_commands.tex
\usepackage{amsmath,amsfonts,bm}

\def\eqref#1{equation~\ref{#1}}
\def\1{\bm{1}}

\DeclareMathAlphabet{\mathsfit}{\encodingdefault}{\sfdefault}{m}{sl}
\SetMathAlphabet{\mathsfit}{bold}{\encodingdefault}{\sfdefault}{bx}{n}

\DeclareMathOperator{\sigm}{sigmoid}
\DeclareMathOperator{\clip}{clip}